\documentclass[a4paper,12pt]{article}
\usepackage{latexsym}
\usepackage{amsmath}
\usepackage{amsfonts}
\usepackage{amssymb}
\usepackage{graphicx}
\usepackage{color}
\usepackage[LGR,T1]{fontenc}

\begin{document}

\title {Hyperbolic angular statistics for globally coupled phase oscillators}

\author{$\,$\\M.-O. Hongler (STI/IMT/LPM EPFL - Lausanne - Switzerland) \\$\,$\\ R. Filliger (Bern Applied Univ. - CH-2500 Biel - Switzerland)
\\ $\,$ \\ Ph. Blanchard (Fak. f\"ur Physik - Uni Bielefeld - Germany)}

 \maketitle

\abstract $\,$

\noindent We analytically discuss a multiplicative noise
generalization  of the  Kuramoto-Sakaguchi dynamics for an
assembly of globally coupled phase oscillators. In the mean field
limit, the resulting class of invariant measures coincides with a
generalized, two parameter family of angular von Mises
probability distributions which is governed by the exit law from
the unit disc of a hyperbolic drifted Brownian motion. Our
dynamics offers  a simple yet analytically tractable
generalization of Kuramoto-Sakaguchi dynamics with two control
parameters. We derive an exact and  very compact relation between
the two control parameters at the onset of phase oscillators
synchronization.

\vspace{0.4cm}

\vspace{0.4cm}

\noindent {\bf PACS number(s)}: 05.45 Xt - 05.10.Gg - 02.50 Ey
\section{Introduction and model}

\noindent The modeling efforts aiming to analytically explain the
phenomenon of collective synchronization of phase oscillators
are, to a large extent, based on the Kuramoto model
\cite{Kuramoto87,sk87} and its mathematical tractable extensions
\cite{SAKA88}. The success of this "toy model" is thoroughly
documented in the recent review by Acebron et al.
\cite{Acebron2005}.

\noindent Despite a very large body of available literature on
coupled phase oscillators, we feel that intimate connections with
directional statistics, typically describing distributions on the
circle, yet remained relatively  neglected. The main purpose of
this note is precisely to highlight this intimate connection and
explore the fruitful consequences that can be derived from it. As
a matter of fact, the mean field stationary solution to the noisy
Kuramoto model proposed by Sakaguchi, (referred from now on as
the KS model) does on the circle and despite to an apparent
non-linearity in the drift term, effectively play  the role the
Ornstein-Uhlenbeck process plays on the line (see discussion in
Mardia \cite{Mardia75}). This remark enables to better
comprehend  why ``this little wonder'', as the KS model is called
in \cite{Acebron2005} and  \cite{Strogatz}, is analytically
tractable. This also suggests that a ''non-flat'' extension of
the mentioned circular Ornstein-Uhenbeck process could be
constructed and used to
 guide us to what  we shall call
here an  hyperbolic extension of the KS model.

\noindent Our starting point is a generalized version of the
Kuramoto model introduced by Sakaguchi \cite{SAKA88}. The model
consists of $N$ non-linearly coupled phase oscillators having
identical natural frequencies $\omega$ and which are perturbed by
{\it multiplicative} Gaussian White Noise. The phase of the
$i$-th oscillator, denoted by $\theta_i$, evolves in time
according to
\begin{equation}
\label{MULTIPLOCOS} \dot{\theta}_i(t) =\omega+{K \over N} \sum
_{j=1}^{N} \sin(\theta_j-\theta_i) +{1 \over N}  \sum
_{j=1}^{N}\left\{\sqrt{\left[1 + C\cos(\theta_j
-\theta_i)\right]} \right\}\xi_i(t).
\end{equation}
\noindent The first summation over the oscillation population in
eq.(\ref{MULTIPLOCOS}) is a
deterministic, non-linear coupling with strength $K>0$. The second
summation in eq.(\ref{MULTIPLOCOS}) couples the oscillators
to random fluctuations through the parameter $0<C\leq 1$. The
resulting state-dependent noise strength modulates the amplitude
of a Gaussian White
Noise (WGN) source $\xi_{i}$'s with constant noise strength $\sqrt{2T}$ i.e.,
\begin{equation}
\label{WNoise} \langle\xi_{i}(t)\rangle =0\;\; \textrm{and}\;\; \langle\xi_{i}(t)\xi_j(t')\rangle
=2T \delta_{ij}\delta(t-t').
\end{equation}

\noindent The additional coupling parameter $C$ controls the
multiplicative character of the noise source. Indeed, for $C=0$
the additive WGN case is reproduced and eq.(\ref{MULTIPLOCOS})
therefore coincides with the original Kuramoto-Sakaguchi model.

\noindent For $C>0$,  we are in presence of multiplicative WGN and
the underlying stochastic integral will be interpreted here in the
It\^o sense.

\noindent Observe that the multiplicative noise in
eq.(\ref{MULTIPLOCOS}) itself depends on interactions  between
the oscillators of the assembly. Several multiplicative noise
sources for KS dynamics have been previously introduced. In
\cite{Reimann99}, the authors discuss a  multiplicative noise KS
in which the WGN modulation affecting  each oscillator is
determined by its dynamical state only, ({\it local noise
modulation}). A global noise modulation exhibiting similarities
with our present model is discussed in \cite{Kim} where the noise
source perturbs  the coupling control parameter itself. The
\cite{Kim} dynamics differs from our present model by the fact
that in the synchronized states, (i.e. when $\theta_{i} =
\theta_{k}$, $ \forall i,k$) the dynamics become deterministic --
a specific feature  absent in our model.

\noindent In general multiplicative noise sources are known to
give rise to new emerging  behaviors, the core mechanism for noise
induced phase transitions \cite{Horsthemke06}. In  the sequel, we
shall indeed observe that the onset of synchronized behaviour
will be strongly modified when $C>0$.

\section{Self-consistent steady solutions}

\noindent  As it is usual, we now analyze eqs. (\ref{MULTIPLOCOS})
and (\ref{WNoise}) in the mean field approximation reached  in
the thermodynamic limit $N\rightarrow\infty$. By going into a
rotating frame,  we can get rid of the common eigenfrequency and
hence, we take from now on $\omega=0$. The collection of
oscillators can then be characterized
 by a circular probability density $\rho(\theta,t)$
where $\rho(\theta+d\theta,t)-\rho(\theta,t)$ stands for  the fraction
of oscillators on the unit circle which, at time $t$,  are located at the
angular position $\theta$. The circular probability
density satisfies the non-linear Fokker-Planck equation
\begin{equation}
\label{nlfp} \frac{\partial \rho}{\partial t} =
-\frac{\partial}{\partial \theta}[RK
\sin(\psi-\theta) \rho]+T\frac{\partial^2
}{\partial \theta^2}[(1+RC\cos(\psi-\theta))\rho],
\end{equation}
with periodic boundary conditions $\rho(\theta+2\pi,t)=\rho(\theta,t)$ and
normalization $\int_{0}^{2\pi}\rho(\theta,t)d\theta=1.$

\vspace{0.3cm}
\noindent In eq.(\ref{nlfp}), the quantity
$R=R(t)\in [0,1]$ yields a measure for the phase
coherence of the oscillators and $\psi=\psi(t)$ represents the average
phase of the assembly. Both $R(t)$ and $\psi(t)$ are used to construct the
complex order parameter
\begin{equation}
\label{ORDERP} R(t) \exp\left(i \psi(t) \right)
:=\int_{-\pi}^{\pi} e^{i\theta}\rho(\theta,t)\,d\theta.
\end{equation}
Following the original procedure used for the KS model in
\cite{Kuramoto87}, we use $R$ to monitor the cooperative state of
the oscillators assembly. When $R=1$ fully synchronized motion is
obtained. The state  $R=0$ characterizes the fully incoherent
behavior. Intermediate states where $0<R<1$ indicates that the
assembly is in a partially synchronized state.

\noindent Under steady state conditions, a self-consistency
condition fixes compatible values for $R$ and $\psi$. This
ultimately enables to extract the critical coupling strength
$K_c$  leading to the onset of partially synchronized dynamical
states.

\vspace{3mm} \noindent For the well known additive noise case
$C=0$ the stationary solution to eq. (\ref{nlfp}) is given by the
famous {\it von Mises angular distribution}
\begin{equation}
\label{KUR} \rho(\theta)= \frac{e^{k\cos(\psi-\theta)}}{2\pi I_0(k)},
\end{equation}

\noindent where  $I_0$ is the modified Bessel function of the first kind
 and where we introduced the dimensionless factor
\begin{equation}
\label{k} k:= R K/T\,.
\end{equation}
Solving self-consistently eq.(\ref{ORDERP})
together with (\ref{KUR}) we find the relation:
  \begin{eqnarray}
\label{SELFCONS} R&=& {I_{1} (k ) / I_{0} (k )},
\end{eqnarray}

 \noindent with $I_1(z)$ the modified Bessel function of the first kind of order $1$.
 For the onset of synchronization, i.e. for
 small but finite $R$, we may expand  Eq.(\ref{SELFCONS})
 to first order in $R$. This yields the well known critical
 coupling $K_c= 2T$ from where
 partially synchronized solutions branch off the fully
 incoherent solution  represented by the uniform
 angular distribution $\rho(\theta)=1/2\pi$.

\noindent Consider now the generalized, multiplicative noise KS
dynamics obtained when  $C>0$ in eq.(\ref{MULTIPLOCOS}). The
stationary solution of eq. (\ref{nlfp}) is given by the so called
hyperbolic von Mises distribution (see \cite{Gruet}, \cite{Jones}
and the discussion in the next section):
\begin{equation}
\label{HypKUR} \rho(\theta)= Z^{-1} \Big[1+\tanh(\eta)\cos(\psi-\theta)\Big]^{-\alpha}
\end{equation}
with
 $$\tanh(\eta):=RC\;\;\;\textrm{and}\;\;\;\alpha:=1-\frac{K}{TC}\,\cdot$$
The normalization $Z$ in eq.(\ref{HypKUR}) involves the Legendre function $P_{-\alpha}^{(0)}$ of order $0$ and degree
$-\alpha$:
$$Z=2\pi P_{-\alpha}^{(0)}(\cosh(\eta))\cosh(\eta)^{\alpha}.$$
In full analogy with  the case $C=0$,  we solve self-consistently eq.(\ref{ORDERP})

 together with (\ref{HypKUR}) and find the relation (see appendix):
  \begin{eqnarray}
\label{SELFCONS2} R&=& \frac{1}{1-\alpha}\,
\frac{P_{-\alpha}^{(1)} (\cosh(\eta))}{
P_{-\alpha}^{(0)}(\cosh(\eta))},
\end{eqnarray}

 \noindent where $P_{-\alpha}^{(1)}$ is the Legendre function of order $1$ and degree $-\alpha$.
 To localize the onset of synchronization,  we again expand  Eq.(\ref{SELFCONS2})
 to first order in $R$, (see appendix) and obtain now the $C$-dependent relation
 \begin{eqnarray}
\label{CC}
 K_c= T(2+C).
 \end{eqnarray}
The remarkably simple form of eq.(\ref{CC}), which is our central
result, relates the coupling $K$ and the multiplicative noise $C$
constants with  the WGN variance $T$. The  presence of the extra
control parameter $C$ in eq.(\ref{CC}) modifies the phase
transition diagram for the synchronization regimes and offers
therefore a natural and very simple generalization  of the
original Kuramoto-Sakaguchi dynamics.

\section{The hyperbolic geometry interpretation of the generalized Kuramoto model}

 \noindent As mentioned before, the KS model stationary
 probability measure coincides with the von Mises distribution and is a cornerstone
distribution in the theory of directional statistics comparable
in many ways with the normal distribution on the line (see Mardia
\cite{Mardia72}). Among the numerous possibilities to obtain  the
von Mises distribution eq.(\ref{KUR}), let us emphasize here that
it describes the exit law from the disc $x^2+y^2\leq 1$ of a
planar Brownian motion starting at the origin subject to a
constant drift vector $\textbf{u}$ of length $k$.

\vspace{3mm}\noindent Recently, a very elegant contribution of
J.-C. Gruet shows  how  a two-parameter hyperbolic von Mises
distribution $\Lambda^{\alpha}(r,du)$ can be constructed from the
exit law of the hyperbolic Brownian motion with drift $\alpha\in
\mathbb{R}$ from the hyperbolic disc of radius $r$, centered at
$i$,  \cite{Gruet}. This hyperbolic extension of the von Mises
distribution can be written as
\begin{equation}
\label{HyperbolicVonMises}   \Lambda^{\alpha}(r,\theta)=
\frac{\big(1+\tanh(r)\cos(\theta)\big)^{-\alpha}}{2 \pi
P_{-\alpha}^{(0)}(\cosh(r))},\;\;\;\theta\in[-\pi,\pi],
\end{equation}
which is obviously  equivalent to the centered (i.e.,  $\psi=0$)
equation (\ref{HypKUR}).  This will establish a connection between
the generalized KS model eq.(\ref{MULTIPLOCOS}) and the hyperbolic
von Mises distribution.

\vspace*{2mm} \noindent Observe that independently of Gruets' work the distribution
eq.(\ref{HyperbolicVonMises}) has been recently introduced by
Jones and Pewsey as a new family of symmetric distributions on
the circle including prominent special cases such as the von
Mises distribution, the wrapped Cauchy distribution and others
\cite{Jones}. This family, call them $h^{\alpha}_{c}(r,\theta)$,
may be described via the $\Lambda^{\alpha}$'s by coupling $r$
with $\alpha$ through the curvature $-c$ of the hyperbolic space
according to the relations $r\rightarrow rc$ and
$\alpha\rightarrow \alpha/c$:
\begin{equation}
\label{JP} h^{\alpha}_{c}(r,\theta) := \Lambda^{\alpha/c}(rc
,\theta), \quad \alpha \in \mathbb{R} \,\,\,{\rm and} \,\, \,c
\in \mathbb{R}^{+}.
\end{equation}
Here $r \geq 0$ stands for a concentration parameter. The index
$-c$ can be interpreted as the curvature of the hyperbolic space.
As the curvature increases to zero, the usual von Mises
distribution is recovered, (see e.g., \cite{Jones} where they
consider the special case $\alpha=1$):
\begin{equation}
\label{JP2} \lim_{-c\nearrow 0}h^{\alpha}_{c}(r,\theta)
=\frac{e^{\alpha r \cos(\theta) }}{2 \pi I_{0}(\alpha r)}.
\end{equation}

\noindent Now it becomes clear,  how to interpret the constant $C>0$ in eq.(\ref{MULTIPLOCOS}). We have
$\eta=RC+ \mathcal{O}((RC)^3)$ and $\alpha=1-K/TC$. Hence, small values of $C>0$ produce
the following representation for the stationary distribution eq.(\ref{HypKUR}):
\begin{equation}
\label{connection}\rho(\theta)= h^{K/T}_{C}(R,\theta).
 \end{equation}
 We then conclude that $-C$ may indeed be interpreted as the curvature of the non-flat manifold
 underlying the generalized  KS model  given in
 eq.(\ref{MULTIPLOCOS}). Note that the hyperbolic model converges for $C\rightarrow 0$
 together with the stationary distribution (\ref{HypKUR}) and the critical coupling (\ref{CC})
  to its flat analogue originally discussed in \cite{SAKA88}.

\section{Conclusions}

\noindent It is immediate to realize that the stationary measure
of the Kuramoto-Sakaguchi (KS) dynamics coincides with the famous
von Mises directional statistics. This distribution on the circle
 can itself be interpreted as the exit law from a
 disc of a drifted Brownian motion. An hyperbolic extension,
  i.e. the exit law of an hyperbolic Brownian motion from a
  Poincar\'e  disk produces a generalized, two-parameter
  von Mises probability distribution which  has very recently
  been considered both from a stochastic and a purely
  statistical point of views. The additional control parameter
  is intimately connected to the curvature of the
  underlying non-flat manifold. By analyzing the
  implication of these features  for the KS dynamics,
  we have been able to construct a simple and natural
  extension of the original KS model for which, to paraphrase
  \cite{Acebron2005} and \cite{Strogatz}, the
  "little wonder"  remains possible. Our  generalization includes,
  besides the coupling strength between the phase oscillators,
  a second control parameter, (i.e. the underlying curvature of
  the curved manifold),  which modulates a multiplicative noise
  source.  Ultimately and in  full analogy with  the original
  KS model, it is straightforward to derive an exact generalized
  synchronization diagram which now depends on two control parameters.

\subsection*{Acknowledgements}

\noindent The authors would like to thank partial support from
the Centro de Ciencas Mathenaicas (CCM)  - Madeira  under the project FCT, POCTI-219.


\section{Appendix}

\noindent An efficient way to derive eq.(\ref{SELFCONS}) is to use the following series representation for the
hyperbolic von Mises distribution valid for $\alpha<1$ \cite{Gruet}:
\begin{equation}
\label{HyperbolicVonMisesSeries}   \rho(\theta)=
\frac{1}{2 \pi}+\frac{1}{\pi}\sum_{m=1}^{\infty}
\frac{\Gamma(1-\alpha)}{\Gamma(m+1-\alpha)}\frac{P_{-\alpha}^{(m)}(\cosh(r))}{
P_{-\alpha}^{(0)}(\cosh(r)) }\cos(m(\psi-\theta)).
\end{equation}
Replacing the above series in eq.(\ref{ORDERP}) and using a trigonometric identity together
 with an orthogonality argument for the cosine functions, we find
trough elementary integration of $\cos(\theta)$ and $\cos(\theta)^2$ over $[-\pi,\pi]$ the equation
  \begin{eqnarray}
Re^{i\psi}&=& \frac{1}{1-\alpha}\frac{P_{-\alpha}^{(1)} (\cosh(\eta))}{ P_{-\alpha}^{(0)}(\cosh(\eta))}e^{i\psi}.
\end{eqnarray}
which is eq.(\ref{SELFCONS}).

\noindent Eq.(\ref{CC}) follows from the above by carefully expanding the Legendre functions $P_{-\alpha}^{(0)}$ and $P_{-\alpha}^{(1)}$ to first order in $R$.
Recall that $\tanh(\eta)=RC$ and hence $\eta=\frac{1}{2}\ln\big(\frac{1+RC}{1-RC}\big).$ Using in the following the
tables in Abramowitz and Stegun \cite{Abramowitz} we have with 8.11.1:
 \begin{eqnarray}
\label{P0}
 P_{-\alpha}^{(0)}(\cosh(\eta))=e^{(\alpha-1)\eta}F\Big(\frac{1}{2},1-\alpha;1;1-e^{-2\eta}\Big)
 \end{eqnarray}
 For $P_{-\alpha}^{(1)}$ more care is needed because of a singularity in the Gauss series for the hypergeometric functions $F$. Using 8.11.1
 we have
 \begin{eqnarray}
\label{P1}
 P_{-\alpha}^{(1)}(\cosh(\eta))=\frac{4e^{(\alpha-1)\eta}}{1-e^{-2\eta}}\lim_{\mu\rightarrow 1}
 \frac{1}{\Gamma(1-\mu)}F\Big(-\frac{1}{2},-\alpha;1-2\mu;1-e^{-2\eta}\Big)
 \end{eqnarray}
 The limit is taken using the fact that (see reflection and duplication formula for $\Gamma$-function
 6.1.17 and 6.1.18):
 \begin{eqnarray}
\label{G}
 \lim_{\mu\rightarrow 1}
 \frac{\Gamma(1-2\mu)}{\Gamma(1-\mu)}=\lim_{\mu\rightarrow 1}
 \frac{2^{2(1-\mu)-1/2}}{\sqrt{2\pi}(1-2\mu)}\Gamma(1-\mu+\frac{1}{2})=-\frac{1}{2}.
 \end{eqnarray}
 We then have with
 15.1.2 in \cite{Abramowitz}:
 \begin{eqnarray}
\label{P1b}
 P_{-\alpha}^{(1)}(\cosh(\eta))=\frac{4e^{(\alpha-1)\eta}}{1-e^{-2\eta}}
 \frac{\alpha(\alpha-1)}{16}\Big(1-e^{-2\eta}\Big)^2 F\Big(\frac{3}{2},2-\alpha;3;1-e^{-2\eta}\Big).
 \end{eqnarray}
 Putting eqs.(\ref{P0}) and (\ref{P1b}) into eq.(\ref{SELFCONS}) and noting that $1-e^{-2\eta}=\frac{2RC}{1+RC}$ we find
in first order in $R$
\begin{equation}
R=-\frac{\alpha}{4}\frac{2RC}{1+RC}=-\frac{\big(1-\frac{K}{TC}\big)}{4}\frac{2RC}{1+RC}
\end{equation}
which reduces for small but finite $R$ to $K=T(2+C).$

\end{document}